\documentclass[%
 reprint,
 superscriptaddress,
 amsmath,amssymb,
 aps,
]{revtex4-2}
\usepackage{graphicx}% Include figure files
\usepackage{dcolumn}% Align table columns on decimal point
\usepackage{bm}% bold math
\usepackage{siunitx}
\usepackage{lineno}
\usepackage{xcolor}
\usepackage{comment}
\usepackage{adjustbox}
\usepackage[utf8x]{inputenc}
\usepackage{ucs}
\usepackage{multirow}
\usepackage{appendix}
\usepackage{ulem}

\begin{document}
%\linenumbers
\title{A First Search for Solar $^8B$ Neutrino in the PandaX-4T Experiment using Neutrino-Nucleus Coherent Scattering}

% !TEX root = ../main.

\def\shKeyLab{School of Physics and Astronomy, Shanghai Jiao Tong University, Key Laboratory for Particle Astrophysics and Cosmology (MoE), Shanghai Key Laboratory for Particle Physics and Cosmology, Shanghai 200240, China}
\def\BUAA{School of Physics, Beihang University, Beijing 102206, China}
\def\BUAALab{Beijing Key Laboratory of Advanced Nuclear Materials and Physics, Beihang University, Beijing, 102206, China}
\def\zzu{School of Physics and Microelectronics, Zhengzhou University, Zhengzhou, Henan 450001, China}
\def\USTClab{State Key Laboratory of Particle Detection and Electronics, University of Science and Technology of China, Hefei 230026, China}
\def\USTCdep{Department of Modern Physics, University of Science and Technology of China, Hefei 230026, China}
\def\BUAALab{International Research Center for Nuclei and Particles in the Cosmos \& Beijing Key Laboratory of Advanced Nuclear Materials and Physics, Beihang University, Beijing 100191, China}
\def\pku{School of Physics, Peking University, Beijing 100871, China}
\def\YaLongSD{Yalong River Hydropower Development Company, Ltd., 288 Shuanglin Road, Chengdu 610051, China}
\def\IAP{Shanghai Institute of Applied Physics, Chinese Academy of Sciences, 201800 Shanghai, China}
\def\CHEPpku{Center for High Energy Physics, Peking University, Beijing 100871, China}
\def\SDUdep{Research Center for Particle Science and Technology, Institute of Frontier and Interdisciplinary Science, Shandong University, Qingdao 266237, Shandong, China}
\def\SDUlab{Key Laboratory of Particle Physics and Particle Irradiation of Ministry of Education, Shandong University, Qingdao 266237, Shandong, China}
\def\UMD{Department of Physics, University of Maryland, College Park, Maryland 20742, USA}
\def\TDLee{Tsung-Dao Lee Institute, Shanghai Jiao Tong University, Shanghai, 200240, China}
\def\MESJTU{School of Mechanical Engineering, Shanghai Jiao Tong University, Shanghai 200240, China}
\def\SYU{School of Physics, Sun Yat-Sen University, Guangzhou 510275, China}
\def\SYUSFI{Sino-French Institute of Nuclear Engineering and Technology, Sun Yat-Sen University, Zhuhai, 519082, China}
\def\NKU{School of Physics, Nankai University, Tianjin 300071, China}
\def\FDU{Key Laboratory of Nuclear Physics and Ion-beam Application (MOE), Institute of Modern Physics, Fudan University, Shanghai 200433, China}
\def\USST{School of Medical Instrument and Food Engineering, University of Shanghai for Science and Technology, Shanghai 200093, China}
\def\SJTUSC{Shanghai Jiao Tong University Sichuan Research Institute, Chengdu 610213, China}
\def\SPEIT{SJTU Paris Elite Institute of Technology, Shanghai Jiao Tong University, Shanghai, 200240, China}

\author{Wenbo Ma}\affiliation{\shKeyLab}
\author{Abdusalam Abdukerim}\affiliation{\shKeyLab}
\author{Chen Cheng}\affiliation{\SYU}
\author{Zihao Bo}\affiliation{\shKeyLab}
\author{Wei Chen}\affiliation{\shKeyLab}
\author{Xun Chen}\affiliation{\shKeyLab}\affiliation{\SJTUSC}
\author{Yunhua Chen}\affiliation{\YaLongSD}
% \author{Chen Cheng}\affiliation{\SYU}
\author{Zhaokan Cheng}\affiliation{\SYUSFI}
\author{Xiangyi Cui}\affiliation{\TDLee}
\author{Yingjie Fan}\affiliation{\NKU}
\author{Deqing Fang}\affiliation{\FDU}
\author{Changbo Fu}\affiliation{\FDU}
\author{Mengting Fu}\affiliation{\pku}
\author{Lisheng Geng}\affiliation{\BUAA}\affiliation{\BUAALab}\affiliation{\zzu}
\author{Karl Giboni}\affiliation{\shKeyLab}
\author{Linhui Gu}\affiliation{\shKeyLab}
\author{Xuyuan Guo}\affiliation{\YaLongSD}
\author{Chencheng Han}\affiliation{\TDLee} %
\author{Ke Han}\affiliation{\shKeyLab}
\author{Changda He}\affiliation{\shKeyLab}
\author{Jinrong He}\affiliation{\YaLongSD}
\author{Di Huang}\affiliation{\shKeyLab}
\author{Yanlin Huang}\affiliation{\USST}
\author{Zhou Huang}\affiliation{\shKeyLab}
\author{Ruquan Hou}\affiliation{\SJTUSC}
\author{Xiangdong Ji}\affiliation{\UMD}
\author{Yonglin Ju}\affiliation{\MESJTU}
\author{Chenxiang Li}\affiliation{\shKeyLab}
\author{Jiafu Li}\affiliation{\SYU}
\author{Mingchuan Li}\affiliation{\YaLongSD}
\author{Shu Li}\affiliation{\MESJTU}
\author{Shuaijie Li}\affiliation{\TDLee}
\author{Qing Lin}\email[Corresponding author: ]{qinglin@ustc.edu.cn}\affiliation{\USTClab}\affiliation{\USTCdep}
\author{Jianglai Liu}\email[Spokesperson: ]{jianglai.liu@sjtu.edu.cn}\affiliation{\shKeyLab}\affiliation{\TDLee}\affiliation{\SJTUSC}
\author{Xiaoying Lu}\affiliation{\SDUdep}\affiliation{\SDUlab}
\author{Lingyin Luo}\affiliation{\pku}
\author{Yunyang Luo}\affiliation{\USTCdep}
%\author{Wenbo Ma}\affiliation{\shKeyLab}
\author{Yugang Ma}\affiliation{\FDU}
\author{Yajun Mao}\affiliation{\pku}
\author{Nasir Shaheed}\affiliation{\SDUdep}\affiliation{\SDUlab}
\author{Yue Meng}\email[Corresponding author: ]{mengyue@sjtu.edu.cn}\affiliation{\shKeyLab}\affiliation{\SJTUSC}
\author{Xuyang Ning}\affiliation{\shKeyLab}
\author{Ningchun Qi}\affiliation{\YaLongSD}
\author{Zhicheng Qian}\affiliation{\shKeyLab}
\author{Xiangxiang Ren}\affiliation{\SDUdep}\affiliation{\SDUlab}
\author{Changsong Shang}\affiliation{\YaLongSD}
\author{Xiaofeng Shang}\affiliation{\shKeyLab}
\author{Guofang Shen}\affiliation{\BUAA}
\author{Lin Si}\affiliation{\shKeyLab}
\author{Wenliang Sun}\affiliation{\YaLongSD}
\author{Andi Tan}\affiliation{\UMD}
\author{Yi Tao}\affiliation{\shKeyLab}\affiliation{\SJTUSC}
\author{Anqing Wang}\affiliation{\SDUdep}\affiliation{\SDUlab}
\author{Meng Wang}\affiliation{\SDUdep}\affiliation{\SDUlab}
\author{Qiuhong Wang}\affiliation{\FDU}
\author{Shaobo Wang}\affiliation{\shKeyLab}\affiliation{\SPEIT}
\author{Siguang Wang}\affiliation{\pku}
\author{Wei Wang}\affiliation{\SYUSFI}\affiliation{\SYU}
\author{Xiuli Wang}\affiliation{\MESJTU}
\author{Zhou Wang}\affiliation{\shKeyLab}\affiliation{\SJTUSC}\affiliation{\TDLee}
\author{Yuehuan Wei}\affiliation{\SYUSFI}
\author{Mengmeng Wu}\affiliation{\SYU}
\author{Weihao Wu}\affiliation{\shKeyLab}
\author{Jingkai Xia}\affiliation{\shKeyLab}
\author{Mengjiao Xiao}\affiliation{\UMD}
\author{Xiang Xiao}\affiliation{\SYU}
\author{Pengwei Xie}\affiliation{\TDLee}
\author{Binbin Yan}\affiliation{\shKeyLab}
\author{Xiyu Yan}\affiliation{\USST}
\author{Jijun Yang}\affiliation{\shKeyLab}
\author{Yong Yang}\affiliation{\shKeyLab}
\author{Chunxu Yu}\affiliation{\NKU}
\author{Jumin Yuan}\affiliation{\SDUdep}\affiliation{\SDUlab}
\author{Ying Yuan}\affiliation{\shKeyLab}
\author{Zhe Yuan}\affiliation{\FDU} %
\author{Xinning Zeng}\affiliation{\shKeyLab}
\author{Dan Zhang}\affiliation{\UMD}
\author{Minzhen Zhang}\affiliation{\shKeyLab}
\author{Peng Zhang}\affiliation{\YaLongSD}
\author{Shibo Zhang}\affiliation{\shKeyLab}
\author{Shu Zhang}\affiliation{\SYU}
\author{Tao Zhang}\affiliation{\shKeyLab}
\author{Yingxin Zhang}\affiliation{\SDUdep}\affiliation{\SDUlab} %
\author{Yuanyuan Zhang}\affiliation{\TDLee}
\author{Li Zhao}\affiliation{\shKeyLab}
\author{Qibin Zheng}\affiliation{\USST}
\author{Jifang Zhou}\affiliation{\YaLongSD}
\author{Ning Zhou}\affiliation{\shKeyLab}
\author{Xiaopeng Zhou}\affiliation{\BUAA}
\author{Yong Zhou}\affiliation{\YaLongSD}
\author{Yubo Zhou}\affiliation{\shKeyLab}

\collaboration{PandaX Collaboration}
\noaffiliation

\date{\today}% It is always \today, today,

\begin{abstract}
A search for interactions from solar $^8$B neutrinos elastically scattering off xenon nuclei using PandaX-4T commissioning data is reported. 
The energy threshold of this search is further lowered compared with the previous search for dark matter, with various techniques utilized to suppress the background that emerges from data with the lowered threshold.
A blind analysis is performed on the data with an effective exposure of 0.48\,tonne$\cdot$year, and no significant excess of events is observed.
Among results obtained using the neutrino-nucleus coherent scattering, our results give the best constraint on the solar $^8$B neutrino flux.
We further provide a more stringent limit on the cross section between dark matter and nucleon in the mass range from 3 to 9\,GeV/c$^2$.
\end{abstract}

\maketitle

% quick functions
\newcommand{\mwba}[1]{\textcolor{violet}{#1}}
\newcommand{\mwbd}[1]{\textcolor{violet}{\sout{#1}}}

% \noindent
% {\color{red}{$^8B$ Neutrino Importance}}

Due to complex fusion processes inside the Sun, neutrinos are continuously generated in large amount. 
As liquid xenon (LXe) detectors dedicated to dark matter (DM) direct search~\cite{aprile2020projected, akerib2020projected, zhang2019dark} have been developed into the multi-tonne scale in recent years, they are now able to reach the sensitivity to detect solar neutrinos via coherent elastic nuclear scattering (CE$\nu$NS).
Among all sources of solar neutrinos, neutrinos produced in the $\beta$ decay of $^8$B are the most likely ones to be detected due to the 15\,MeV Q value.
The flux of $^8$B solar neutrinos on Earth has been measured to be approximately 5$\times$10$^6$\,cm$^{-2}$s$^{-1}$~\cite{agostini2018comprehensive, aharmim2013combined}, and its CE$\nu$NS has an energy spectrum hardly distinguishable from that of a 6\,GeV/c$^2$ DM particle in LXe.
No experimental determination of the solar neutrino flux using its CE$\nu$NS signal has been made yet.
Recently, the XENON1T collaboration has published a search for the $^8$B CE$\nu$NS signal using 0.6 tonne$\cdot$year data with no excess found~\cite{aprile2021search}. 
Due to the low nuclear recoil (NR) energy from the $^8$B CE$\nu$NS, it is crucial to lower the energy threshold. 
In this letter, we report a search for CE$\nu$NS induced by the solar $^8$B neutrinos using the commissioning data of PandaX-4T (Run0) based on a blind analysis, with a dedicated data selection,  which lowered the energy threshold (defined as the energy having signal acceptance of 1\%) from 1.33 to 0.95 keV.

PandaX-4T dark matter direct search experiment is located in China Jinping underground Laboratory (CJPL)~\cite{kang2010status, li2015second}.
the PandaX-4T experiment utilizes a dual-phase xenon time projection chamber (TPC) with a sensitive volume of 3.7\,tonne of LXe, and two arrays of photo-multipliers (PMTs) on the top and bottom of the TPC, consisting of 169 and 199 Hamamatsu 3-inch R11410-23 PMTs, respectively.
Both the primary scintillation ($S1$) and the delayed proportional scintillation from drifted electrons ($S2$) of an event are collected by the PMTs, allowing 3-D position reconstruction with a resolution of about a few millimeter for $S2$s of $\sim$100\,photoelectron (PE) on the longitudinal and transverse directions, based on the time difference between the $S1$ and $S2$, and the PMT pattern of the $S2$, respectively.
The waveforms of the PMTs are digitized by CAEN V1725 digitizers and read out under the self-trigger mode when the pulse amplitude is approximately 1/3\,PE above the baseline~\cite{he2021500}.  
More details of the detector apparatus can be found in Refs.~\cite{meng2021dark, zhao2021cryogenics, he2021500}.
PandaX-4T has reported the most stringent constraint on the spin-independent cross sections between the nucleon and DM with the DM mass from 5\,GeV/c$^2$ to 10\,TeV/c$^2$~\cite{meng2021dark} using the 0.63-tonne-year data from Run0.

% Updated Data selection

Compared with the search reported in Ref.~\cite{meng2021dark}, new data selections are developed to enhance the detection efficiency and to minimize the extra background that emerged from data.
Thresholds of the $S1$ and $S2$ are lowered to 0.3\,PE and 65\,PE (both in charge), respectively, as compared with the 2\,PE and 80\,PE in Ref.~\cite{meng2021dark}.
The systematics of the background and the energy reconstruction at such low threshold form the core of this analysis.
With these thresholds, two sets of data used in Ref.~\cite{meng2021dark} with a total live time of about 7.5 days show a higher noise rate, likely due to micro-discharging in the TPC, and are removed from this analysis.
The data selection cuts used in this analysis are described as follows.
We adopt four selection cuts from the previous analysis~\cite{meng2021dark}, the diffusion cut ($S2$ widths compatible with the expected fluctuation on the electron arrival time), the veto PMT cut (no signal in the PMTs outside the field cage), the fiducial volume cut (FV, 2.67\,tonnes), and the single scatter cut (only one $S2$ above 50\,PE in the 1-ms event window).
Events with large signals are observed to be followed by small afterglow signals in PandaX-4T and other experiments~\cite{aprile2014observation, akerib2020investigation}.
These afterglow signals usually are single electrons (SEs) which have a strong correlation with the previous large $S2$ in both time and position.
% Fig.~\ref{fig:signal_vs_deltaT} shows the correlation between small $S2$s and previous large $S2$s in time and position.
Compared with Ref.~\cite{meng2021dark}, a more stringent afterglow veto based on the time and position difference to the previous event is implemented.
Events with a time difference to previous $S2$ ($>$2000\,PE) less than 50\,ms or position difference smaller than 100\,mm are excluded.
In addition, we veto the event unless the total charge per unit time and the number of $S1$s in the preceding 1-ms window have returned back to normal.
% These two afterglow cuts reduce the effective live time to 64.7 days.
% \textcolor{red}{The cut off after a > 10000 PE signal is enhanced from 22~ms to 50~ms to further reject potential accidentally occurred multiple-electron S2s.
% Moreover, the averaged charge-density evolution for subsequent signals following various kinds of energetic signals are studied and parameterized to predict the unquiet window after energetic signals.
% And time-position-correlation cut is made that for each >~2000~PE major S2, within 100~mm and 50~ms subsequent $S2$s are rejected.}
% \sout{We require that in each millisecond time window the total charge is less than 14\,PE and the number of $S1$-like ($<$2\,PE) hits less than 15.}
% \textcolor{red}{We require that for each event, the remnant charge density due to previous large S2 is no more than 14\,PE per millisecond, and number of $S1$-like ($<$2\,PE) signals in the 1-ms window before the event is less than 15.}
The afterglow veto cut also includes a set of ``activity'' requirements on an event waveform, that the ratio between the main $S2$ charge and the total event charge $\mathcal{F}_{S2}$=$q_{S2}/q_{\textrm{event}} > 5/6-150/q_{\textrm{event}}$, 
the integrated charge in the preceding event window to be less than 20\,PE, 
and the main $S1$ to be the only signal within 4-$\mu$s around it.
The effective live time of this analysis is estimated to be 64.7\,days.

The signal expectation in this analysis is produced by a two-step simulation.
The first step is the same as in Ref.~\cite{meng2021dark}, in which the correlated distribution in $S1$ and $S2$ are produced according to a fit to the calibration data, later referred to as the signal model.
In the second step, a dedicated waveform simulation (WS) is developed. 
% A dedicated simulation of the signal waveform (WS) is developed as a critical tool for this analysis.
% A simulation of the signal waveform (WS) is developed to assess the signal reconstruction efficiency.
The waveform of the $S1$ is assembled using sampled $S1$ hits from the neutron calibration data, similar to the procedure in Ref.~\cite{akerib2021simulations}.
The waveform of the $S2$ at any given position is assembled using individual SE waveforms from the data, with the reconstructed position within a 40-mm radius circle.
The width of the overall assembled waveform at a given depth in the TPC is required to satisfy the diffusion relation observed from the data.
Effects of PMT afterpulsing, delayed electrons~\cite{sorensen2017two, sorensen2017electron,akerib2020investigation,akerib2021improving}, and photo-ionization of impurities after a large $S2$ are implemented in the WS according to the data.
More details can be found in the appendix.

\begin{figure}[htp]
    \centering
    \includegraphics[width=0.5\textwidth]{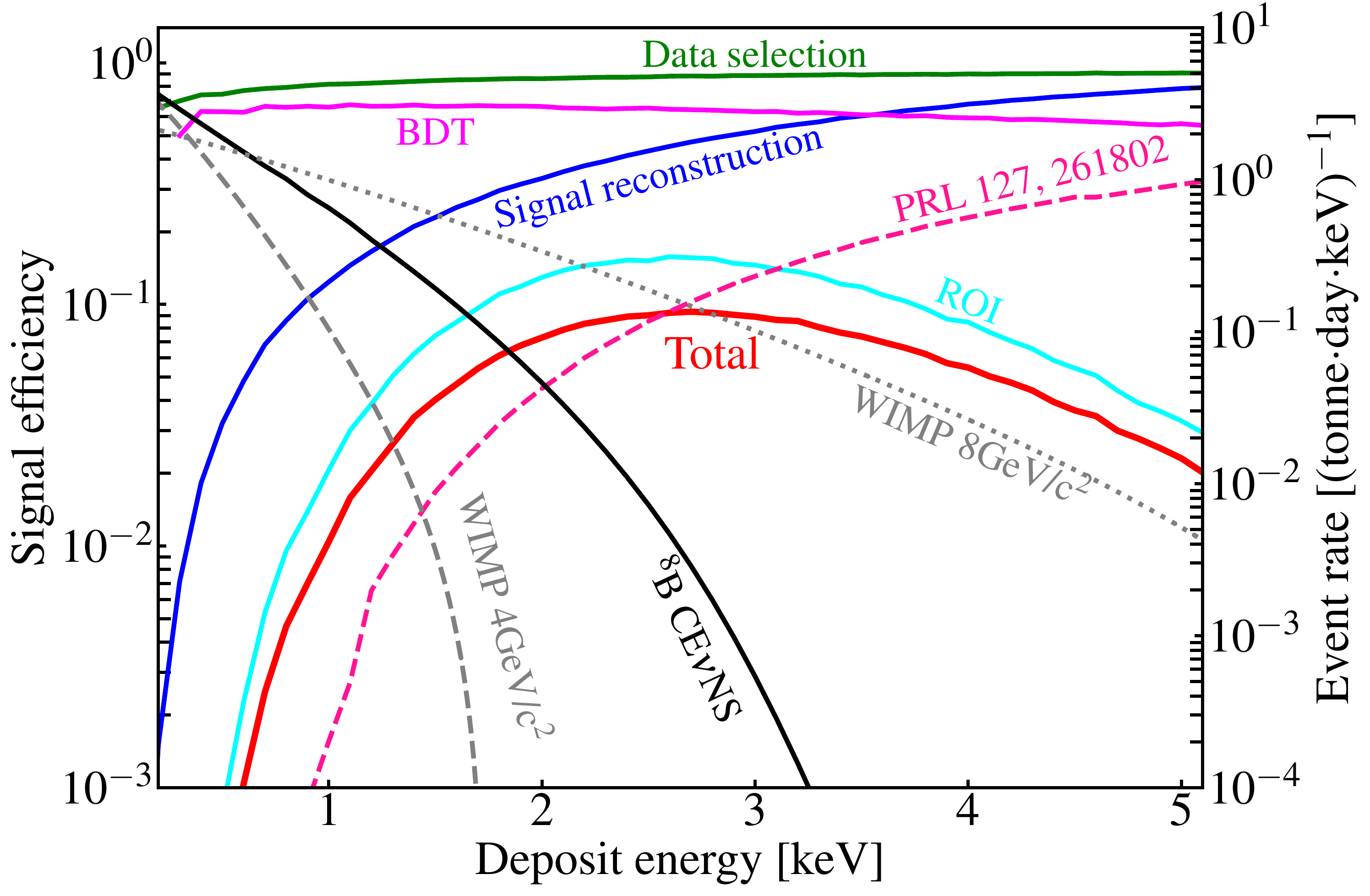}
    \caption{
    Total efficiency (red solid line) to the solar $^8$B neutrino CE$\nu$NS in this analysis with the number of $S1$ hits to be 2 or 3.
    The blue, green, cyan, and magenta solid lines represent the signal efficiencies due to the signal reconstruction, data selection, ROI, and BDT, respectively.
    The signal efficiency in the previous study~\cite{meng2021dark} is also given in the pink dashed line as a reference.
    The ideal spectra of the solar $^8$B CE$\nu$NS and the DM-nucleus interaction with the DM mass of 4 (8)\,GeV/c$^2$ with an assumed DM-nucleon cross section of 10$^{-44}$\,cm$^2$ are overlaid as well in the black solid and grey dashed (dotted) curves, respectively, with the scale indicated on the right axis.}
    \label{fig:eff}
\end{figure}

% clustering, classification and pairing efficiencies
The total efficiency to the $^8$B CE$\nu$NS consists of four components (see Fig.~\ref{fig:eff}): 
1) the signal reconstruction, 
2) the data selections discussed two paragraphs earlier, 
3) the region-of-interest (ROI), and 
4) a cut based on boosted decision tree (BDT, see later text).
the signal reconstruction includes clustering of PMT hits into signal pulses, classification of the signal pulses into $S1$s and $S2$s, and pairing of the classified $S1$s and $S2$s into incident events.
Each step of the signal reconstruction is affected by the presence of dark noises and stray electrons.
%\textcolor{violet}{The data selection cuts are as mentioned earlier.}
For the ROI, we require the number of coincident PMT hits in an $S1$ to be either 2 or 3 in this analysis.
the events with only a single-hit $S1$ are mostly accidental background originating from the PMT dark noises, and are excluded from the ROI due to a poor signal-to-background ratio. 
%\textcolor{black}{
The $S2$ charge range, uncorrected for spatial dependence, is further optimized to be 
65$-$230\,PE for 2-hit $S1$ and 65$-$190\,PE for 3-hit $S1$ based on the expected signal-to-background ratio.
%and will be described in later text.
This ROI requirement has dominating effects on the signal efficiency.
The efficiencies of 1), 2) and 3) are estimated using the WS and validated
by the neutron calibration data, with their fractional difference (14\%) taken as the systematic uncertainty.

\begin{figure}[htp]
    \centering
    \includegraphics[width=0.42\textwidth]{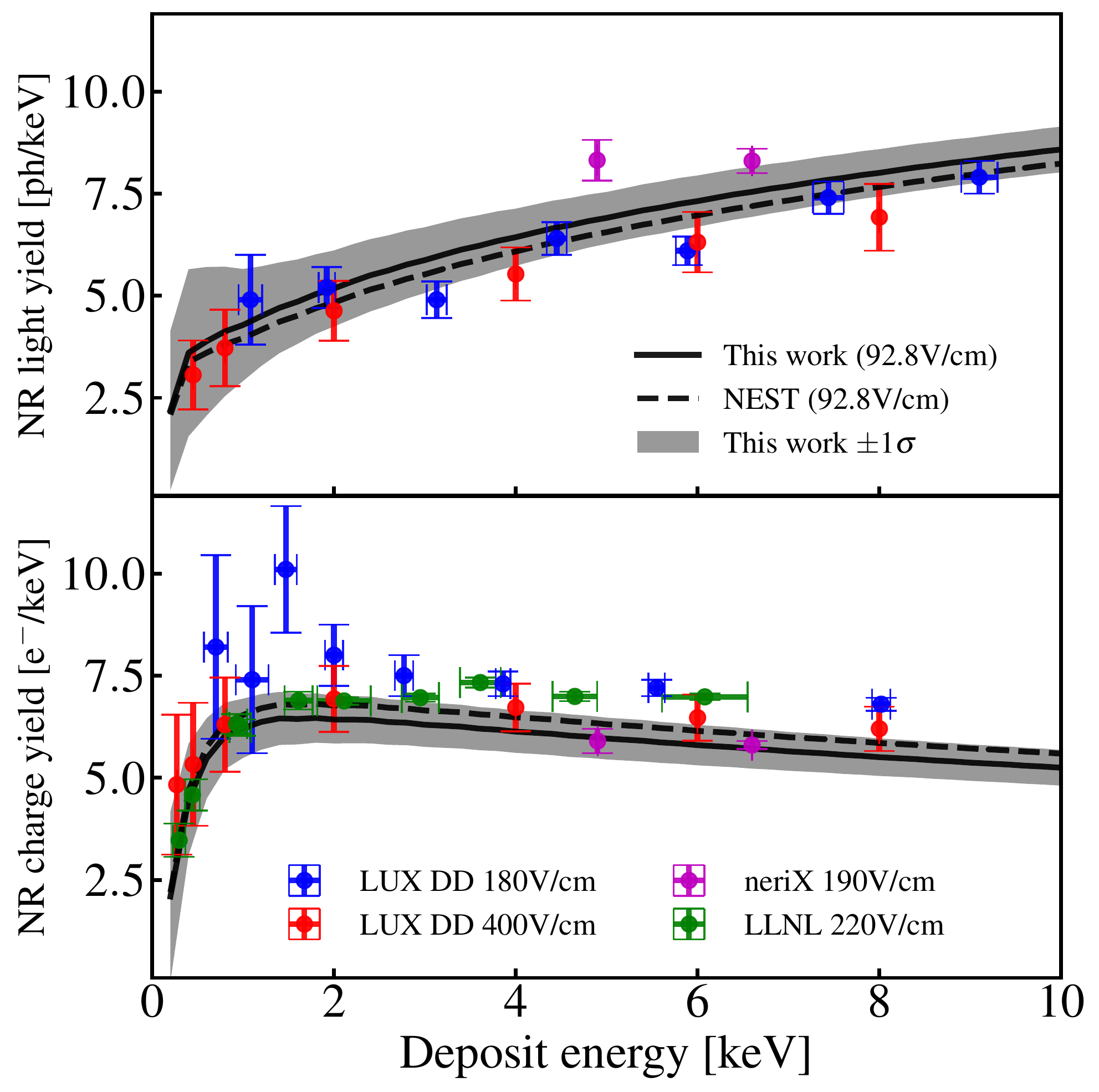}
    \caption{
    Comparison between the light (top panel) and charge (bottom panel) yields used in this analysis (in the solid black lines) with the nominal NEST v2.3.6~\cite{szydagis2018noble} (dashed black lines) and other measurements taken at different drift electric fields~\cite{akerib2016low, huang2020ultra, aprile2018simultaneous, lenardo2019measurement}.
    The 1$\sigma$ uncertainty from Ref.~\cite{szydagis2022noble} for the light and charge yields are shown in the grey bands.
    }
    \label{fig:lycy}
\end{figure}

% Signal model
We take the calculated deposit energy spectrum of the solar $^8$B CE$\nu$NS in LXe from Ref.~\cite{ruppin2014complementarity}, which is shown in Fig.~\ref{fig:eff}. 
% along with deposit energy spectra of the WIMP with the mass of 4 and 8\,GeV/c$^2$.
% The ROI of $S1$ is from xx to xx\,PE and of $S2$ from xx to xx\,PE.
% The hit number is required to be two.
The signal model implements the light and charge production in LXe following the NEST v2.3.6 parametrization~\cite{szydagis2018noble}, and the response of signal detection in the PandaX-4T detector, similar to Ref.~\cite{meng2021dark}. 
The light and charge yields are extrapolated from the one used in Ref.~\cite{meng2021dark}, which has its model parameters fit to the neutron calibration data in the energy region of the DM search (see Fig.~\ref{fig:lycy}).
% The systematic uncertainty of the signal model is dominated by the uncertainty of light and charge yields in LXe, due to the limited measurements of them in such a low energy range.
We adopt the relative uncertainties of the light and charge yields from NEST~\cite{szydagis2022noble}, which is based on a global fit to all available measurements, and conservatively assume them to be uncorrelated.
% The light and charge yields, and their uncertainties, used in this analysis are shown in Fig.~\ref{fig:lycy}.
% Results from several measurements~\cite{akerib2016low, huang2020ultra, aprile2018simultaneous, lenardo2019measurement} and NEST prediction~\cite{szydagis2018noble} under nominal field strength of PandaX-4T are also presented.

% ER and neutron bkg
The background composition is the same as Ref.~\cite{meng2021dark}.
With loosened $S1$ and $S2$ selections, the accidental coincidence (AC) background increases significantly in comparison to Ref.~\cite{meng2021dark}, which dominates the overall background.
% This increases the background rate of accidental coincidence (AC) of isolated $S1$s and $S2$s.
% It makes the ER and NR background from radioactivity in LXe and surrounding detector materials sub-dominated by that from AC.
The electronic recoil (ER), NR, and surface background are estimated using the same method as in Ref.~\cite{meng2021dark} but with the new data selections and the ROI cut.
% The surface background is negligible.
% In the analysis we divided the data into three categories according to the number of hits of $S1$.

\begin{figure*}
    \centering
    \includegraphics[width=0.7\textwidth]{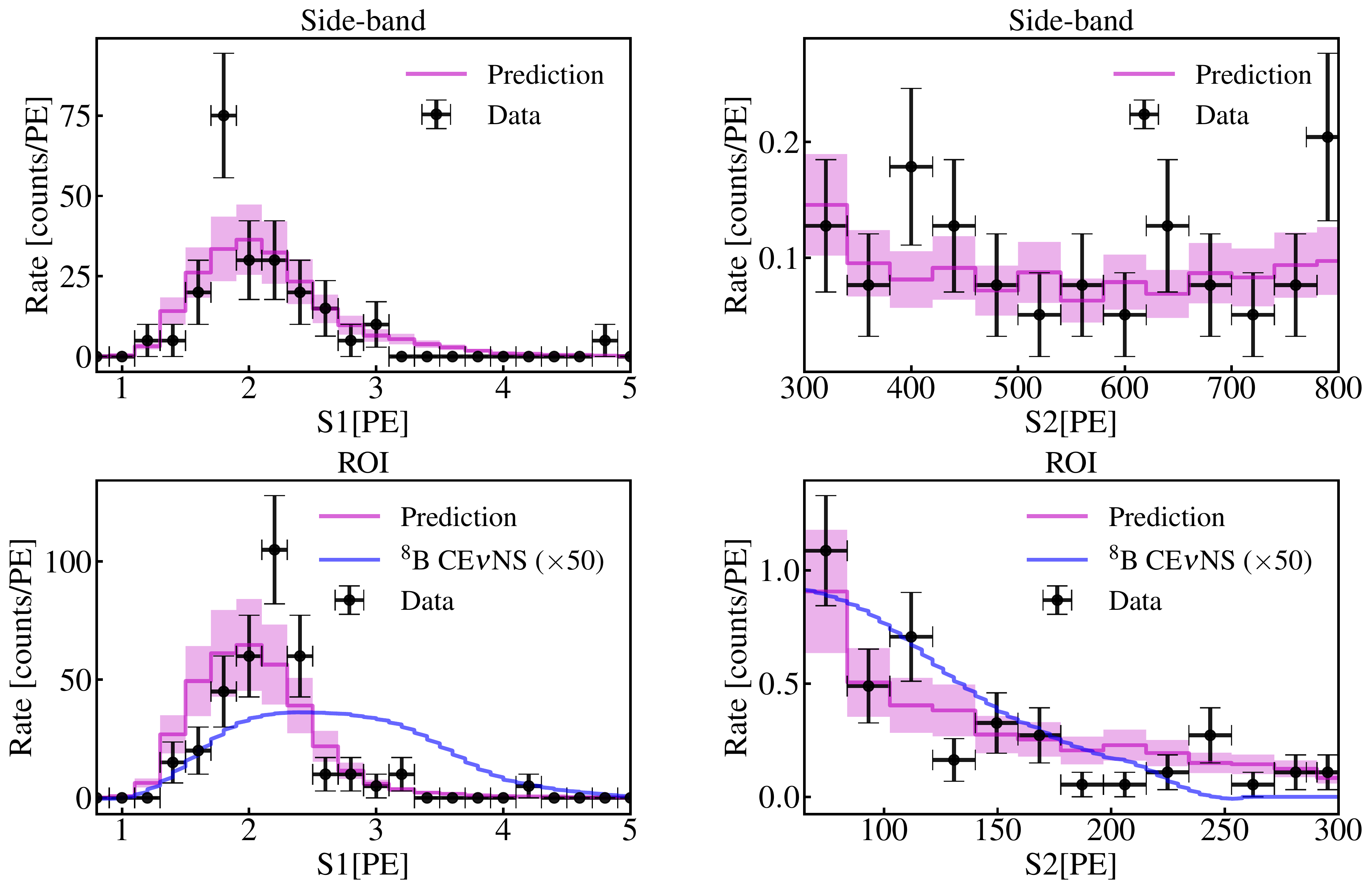}
    \caption{
    The $S1$ (left panels) and $S2$ (right panels) spectra in the side-band (top panels) and ROI (bottom panels) for the 2-hit data, with 
    the data and corresponding predictions overlaid. 
    the shaded regions represent the 1$\sigma$ uncertainty of the prediction (30\%).
    We also overlay the expected $^8$B CE$\nu$NS spectra (scaled up by 50) in the bottom panels, shown in blue solid lines. 
    The goodness-of-fit p-values of the $S1$ and $S2$ spectra in the side-band and ROI are all no less than 0.1.
    }
    \label{fig:ac_validation}
\end{figure*}

% The AC background is constructed from data waveforms.
The rate of the AC background is estimated using random $S1$s and $S2$s identified in the data.
The $S2$s are first selected from a waveform ($\sim$1000 per day within 65 to 300\,PE), then we search backward for 1.5\,ms for a main $S1$. 
The 1.5-ms window is choosen so that the corresponding ``activity'' cuts are sufficiently similar to those mentioned earlier.
% The choice of the 1.5 ms is that our “afterglow” cuts in such a time window are sufficiently close to the ones we used in the physical event window within the TPC drift time. 
The AC pair is formed when the time difference between the $S1$ and $S2$ is within [0.9, 1.5]\,ms, beyond the TPC’s maximum drift time (off-window), to guarantee that there is no correlation.
% \textcolor{red}{
% The rate of uncorrelated $S2$ within 65 to 300\,PE after all data selection cuts is estimated to be about 1000 per day. }
% Alternative event building with a 1.5-ms window is performed, and events with drift time in range of [0.9, 1.5]\,ms provide isolated S2 waveforms.
% S1 and S2 rates are measured from 900-1500\,$\mu$s observation.
% Based on S1 and S2 rates, ROI AC prediction is made for the fiducial volume 40-800\,$\mu$s of real data.
% Furthermore, waveforms containing $S1$s and $S2$s are randomly paired to produce large statistics simulated AC events.
To enlarge the statistics of the AC samples, a ``scrambled'' waveform data set is constructed. 
The waveform of the selected $S2$ is concatenated after a 1-ms segment randomly selected from our recorded data, which on average contains 6.3 (0.01) of the $S1$-like signals with the $S1$ hit equals to (larger than) 1, primarily from dark noises. 
This ``scrambled'' data get passed to the aforementioned software reconstruction and data selection. 
The predicted number of AC events in the ROI in the 2- and 3-hit regions can be found in Table~\ref{tab:bkg}.
The diffusion cut is the most effective cut, which suppresses the AC by a factor of 8 or so.
The AC model is validated using the events with the $S2$ in the range from 300 to 800\,PE (referred to as the side-band data) and within the FV, which is dominated by the AC (see Table~\ref{tab:ac_validation}).
% BDT has great rejection power against AC, so that we assume pre-BDT data are mostly AC.
% The physical background in side-band region is estimated to be mostly negligible, which is shown in Table~\ref{tab:ac_validation}, so that we consider side-band data are mostly AC as well. % Re-write
The comparison between the side-band data and the prediction is given in Table~\ref{tab:ac_validation}, yielding a good agreement.
The comparison between the $S1$ and $S2$ spectra of the prediction and the side-band data for the 2-hit region is shown in Fig.~\ref{fig:ac_validation}.
To be conservative, we take 30\%, which is the difference (error-weighted standard deviation) in the normalized $S2$ spectra, as the systematic uncertainty of the AC model.

\begin{table}[htp]
% \footnotesize
\caption{
ROI comparison: prediction vs. observation in the optimized $S2$ ranges.
Number of the pre- and post-BDT events are listed in separate rows.
The observed events after unblinding are shown in the last column.
}
\begin{tabular}{c|c|c|cccc|c|c|c}
\hline\hline
  \multirow{2}{*}{N$_{\textrm{hit}}$}  & $S2$ range & \multirow{2}{*}{BDT} & \multirow{2}{*}{ER} & \multirow{2}{*}{NR} & \multirow{2}{*}{Surf} & \multirow{2}{*}{AC} & Total & \multirow{2}{*}{$^8$B} & \multirow{2}{*}{\textbf{Obs}} \\
     & [PE] &  &  &  &  &  & BKG &  &  \\
\hline\hline
   \multirow{2}{*}{2}    & \multirow{2}{*}{65-230} & pre & 0.04 & 0.10 & 0.14 & 62.43 & 62.71 & 2.32 & \textbf{59} \\
   & & post & 0.02 & 0.04 & 0.03 & 1.41 & 1.50 & 1.42 & \textbf{1} \\
   \hline
   \multirow{2}{*}{3}    & \multirow{2}{*}{65-190} & pre & 0.01 & 0.05 & 0.08 & 0.79 & 0.93 & 0.42 & \textbf{2}\\
   & & post & 0.00 & 0.02 & 0.03 & 0.02 & 0.07 & 0.29 &  \textbf{0} \\
\hline\hline
\end{tabular}
\label{tab:bkg}
\end{table}

\begin{table}[htp]
\caption{
Side-band comparison: prediction vs. observation for $S2$ within [300, 800]\,PE.
}
\begin{tabular}{c|cc|c|c}
\hline\hline
% \multirow{2}{*}{N$_{\textrm{hit}}$}   & \multicolumn{3}{c}{Side-band} \\
    %  & Phys. & AC & Total & Data \\
N$_{\textrm{hit}}$ & Physical & AC & Total & \textbf{Obs} \\
\hline\hline
1    &  9.4 & 2060.5 & 2069.9 & \textbf{2043} \\
2    & 10.1 &   33.8 &   43.9 & \textbf{47} \\
3    &  6.9 &    2.2 &    9.1 & \textbf{7} \\
\hline\hline
\end{tabular}
\label{tab:ac_validation}
\end{table}

A BDT algorithm~\cite{hoecker2007tmva} is trained to optimize the $^8$B CE$\nu$NS selection against the AC background.
The $S2$s of the AC events are mostly generated out of the fiducial region (such as the surface of electrodes and the gas region), and the $S1$s are mostly dark noises (see Ref.~\cite{abdukerim2022study}), both having different characters from the physical events.
% As a result, a set of variables which have some sensitivities to identify them.
% $S1$s and $S2$s of AC are observed to have sizable difference in the shape and pattern with those of physical events.
The input variables of the BDT concern features related to the charge, width, top-bottom asymmetry, and PMT top patterns of the $S1$ and $S2$ signals.
% The signal and background samples used for training are from simulation and data, respectively.
The training and testing samples of the $^8$B signal in the BDT are from the WS with the ($S1$, $S2$) distribution following our $^8$B signal model.
% The BDT efficiency and rejection power are correlated and adjustable. 
The BDT cut value and the $S2$ range for each $S1$ hit bin are determined by maximizing the probability of discovering a $^8$B signal under our background model, with results summarized in Table~\ref{tab:bkg}.
% with significance of 2 sigma, under a background prediction of $B$ which also takes into account the rest ER and NR background from radioactivity.
% The figure of merit of such probability is defined as
% \begin{equation}
% \sum_{N_{obs}>N_{cri}} \frac{(B+S)^{N_{obs}}}{N_{obs}!}e^{-(B+S)},
% \end{equation} 
% where $S$ is the prediction of $^8$B signal and $N_{cri}$ is defined as the critical number, at which it has a probability of less than 2.28\% (one side of 2 sigma significance) to be observed with background prediction of $B$.
% The optimized $S2$ ranges are shown in Table~\ref{tab:bkg}.
The optimized BDT efficiency of the $^8$B signal is shown in Fig.~\ref{fig:eff}.
The BDT reduces the $^8$B CE$\nu$NS signal (AC background) by about  39\% (98\%) and 31\% (96\%), respectively, for the 2- and 3-hit bins.
%\textcolor{violet}{The test for ranking and correlation of BDT variables were carried out to understand its \uline{drastic} background-rejection power.
%The intrinsic shape of few-photon S1s (dominated by light propagation~\cite{lux_pulse_shape_paper}) and few-electron S2s (dominated by longitudinal diffusion) \uline{degenerate} and smear out far from expectation value than these many-quanta ones.
Most of the rejection power against the AC is gained through the parameters related to the $S2$ waveform shape and its top charge pattern, and we observe almost no correlation in the $S1$ and $S2$ discriminants.
%By \uline{randomly changing} S1 and S2s in the WS signal dataset, the signal efficiency remains unchanged within 1\%, supporting that negligible S1-S2 correlation \uline{inside the discriminator.}}
The uncertainties of the BDT efficiency to the $^8$B CE$\nu$NS and the DM signals are studied using the neutron calibration data. 
To improve the statistics in the ROI, especially for $S2$ less than 100\,PE, the minor $S2$s of the neutron double-scatter events are used.
A difference of 14\% and 13\% are observed for the 2-hit and 3-hit ROI, respectively, taken as the systematic uncertainties.
The systematic uncertainty of the BDT efficiency to the AC background is estimated by checking the performance on an alternative AC model using a more traditional approach based on the random pairing of the isolated $S1$s and $S2$s~\cite{abdukerim2022study}, leading to an uncertainty of 19\% and 18\% in the 2-hit and 3-hit bins.

\begin{table}[htp]
    \centering
    \small
    \caption{
    List of the constrained nuisance parameters that are included in the final statistical interpretation (see text), along with the standard deviations of their Gaussian constraints.
    }
    \begin{tabular}{ll|cc|l}
    \hline\hline
    \multirow{2}{*}{Nuisance parameters} &   & \multicolumn{2}{c|}{Stdev.}   & \multirow{2}{*}{Estimated by} \\
                                         &   & 2-hit & 3-hit                 &                               \\
    \hline
    pre-BDT eff.        & $\delta_{\epsilon}$   & \multicolumn{2}{c|}{0.14}  & \textcolor{black}{WS vs. NR} \\ 
    NR signal rate      & $\delta_s f_i$        & \multicolumn{2}{c|}{$f_i$} & NEST uncert.~\cite{szydagis2022noble} \\
    AC rate             & $\delta_b$            & \multicolumn{2}{c|}{0.30}  & Pred. vs side-band\\
    BDT eff. to signal  & $\delta_{\textrm{BDT},s}^i$ & 0.14 & 0.13 & \textcolor{black}{WS vs. NR} \\
    BDT eff. to AC      & $\delta_{\textrm{BDT},b}^i$ & 0.19 & 0.18 & Alter. models~\cite{abdukerim2022study} \\
    Solar $^8$B flux    & $\delta_{\Phi}$             & \multicolumn{2}{c|}{0.04} & Ref.~\cite{aharmim2013combined} \\
    \hline\hline    
    \end{tabular}
    \label{tab:systematic_uncertainties}
\end{table}

\begin{figure}[htp]
    \centering
    \includegraphics[width=0.48\textwidth]{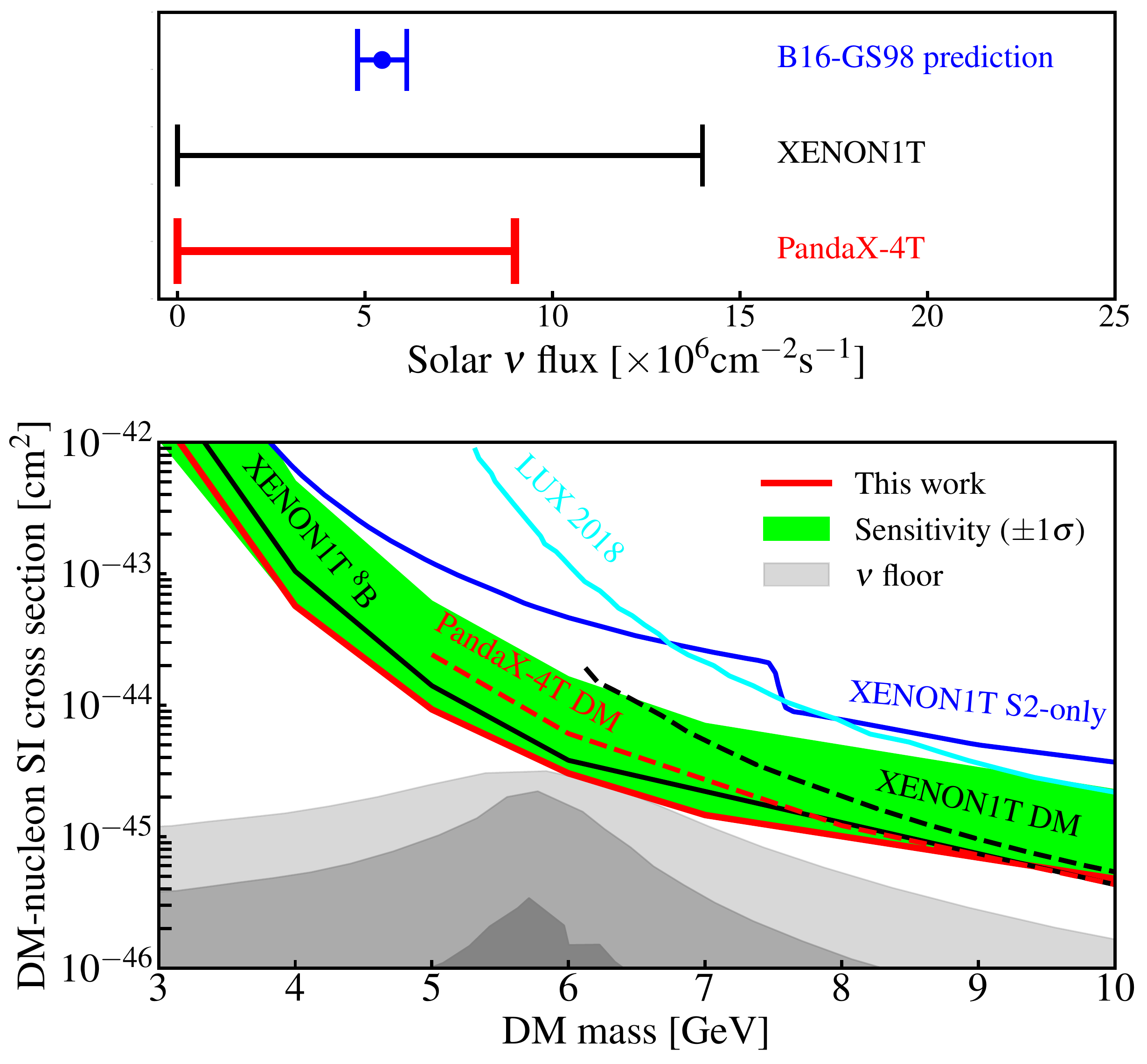}
    \caption{
    Top panel: our constraint on the solar neutrino flux using the CE$\nu$NS analysis, along with the XENON1T results~\cite{aprile2021search} using the same detection channel and the B16-GS98 standard solar model prediction~\cite{vinyoles2017new}.
    Bottom panel: updated constraints on the DM-nucleon spin-independent cross section.
    The red solid and dashed lines represent the PandaX-4T results from this and the previous searches~\cite{meng2021dark}, respectively.
    The black solid and dashed lines represent the results from XENON1T with and without optimization in the low-energy region~\cite{aprile2018dark, aprile2021search}.
    Several results from other experiments~\cite{aprile2019light,akerib2017results} are also shown.
    The neutrino floors (probability for an ideal xenon detector to see less-than-3$\sigma$-significance DM signal)~\cite{ruppin2014complementarity} under different exposure assumptions (1, 10, and 1000 tonne-year from top to bottom) are shown in the grey shaded regions.
    The green region represents the $\pm$1$\sigma$ sensitivity band for the DM search.
    % \textcolor{violet}{(suggest legend in figure: lux wimp 2018, xenon1t s2-only 2019, xenon1t wimp 2018, xenon1t b8 2021, pandax-4t wimp 2021.)}
    }
    \label{fig:results}
\end{figure}
% \noindent
% {\color{red}{Results}}\\

The data within the ROI were blinded before we finalized the data selection, the background and signal models, the ROI, and the BDT optimization.
We then unblinded the data and checked the events before and after applying the BDT.
We show the comparison of the $S1$ and $S2$ spectra between the prediction and data before applying the BDT in Fig.~\ref{fig:ac_validation}.
The observed number of the events in the ROI for the 2- and 3-hit regions are given in Table~\ref{tab:bkg}.
After unblinding, 1 (with $S1$=1.6\,PE and $S2$=165\,PE) and 0 events that survive the BDT are found in the 2- and 3-hit ROI, respectively.

% Since AC is the dominant background in ROI and it has minor difference in $S2$ spectrum compared with solar $^8$B neutrino CE$\nu$NS events, 
We perform a simple statistical interpretation based on 2-bin profile likelihood ratio (PLR) analysis~\cite{baxter2021recommended} using the 2- and 3-hit data.
% The data in 1-hit region is excluded from ROI for interpretation before unblinding due to its high ratio of AC background to solar $^8$B signal, and serve as a cross check for AC model.
The binned likelihood is defined as~\footnote{Penalty term $G(\delta_{\Phi})$ is not used when interpreting results under the signal hypothesis of solar $^8$B CE$\nu$NS.}:
\begin{equation}
    \begin{aligned}
    \mathcal{L} = & G(\delta_{\epsilon}) G(\delta_s) G(\delta_b) G(\delta_\Phi) \\
    & \times \left[ \prod \limits_i G(\delta_{\textrm{BDT},s}^i) G(\delta_{\textrm{BDT},b}^i) \frac{\lambda_i^{N_i}}{N_i !} e^{-\lambda_i} \right] ,
    \label{eq:likelihood}
    \end{aligned}
\end{equation}
% \begin{equation}
%     \mathcal{L} = G(\boldsymbol{\eta}) \prod \limits_i \frac{\lambda_i^{N_i}}{N_i !} e^{-\lambda_i} \cdot G(\boldsymbol{\theta_i}),
%     \label{eq:likelihood}
% \end{equation}
where the index $i$ represents the hit number of $S1$ (2 or 3), and $\boldsymbol{\delta}$ ($\boldsymbol{\delta^i}$) is series of the constrained nuisance parameters, which are correlated (independent) between the 2- and 3-hit bins with a Gaussian penalty $G$ with the mean at zero. 
The set of parameters include $\delta_\epsilon$, $\delta_s$, $\delta_b$, $\delta^i_{\textrm{BDT}, s}$, $\delta^i_{\textrm{BDT},b}$, and $\delta_\Phi$, corresponding to the relative uncertainties of the pre-BDT efficiency (including the signal reconstruction, data selection, and ROI), the NR signal rate, the AC background rate, the BDT cut efficiency to the NR signals, the BDT efficiency to the AC background, and the $^8$B neutrino flux, respectively.
The 1$\sigma$ values of the nuisance parameters are summarized in Table~\ref{tab:systematic_uncertainties}. 
The parameter $\delta_s$ is factored together with the fractional uncertainty of the signal rate $f_i$ which depends on the the signal spectrum ($f^\nu_i$ for the $^8$B CE$\nu$NS signal and $f^\chi_i$ for the DM signal), in order to reflect the common origin of $f_i$.
Typical numbers of $f_i$ are 0.45 (0.60), 0.29 (0.39), and 0.16 (0.24) for 4-GeV/c$^2$ DM, the $^8$B CE$\nu$NS, and 8-GeV/c$^2$ DM in the 2-hit (3-hit) region.
% A single nuisance parameter $\delta_{s}$ reflects the common origin of $f_i$.}
% Uncertainty for other background events is negligible and ignored here.
%All nuisance parameters and their constraints are assumed to be independent of the deposit energy.
%and are listed in Table~\ref{tab:systematic_uncertainties}.
$\lambda_i$ is the expected count while $N_i$ is the observed count. Specifically, under the hypotheses of a) the solar $^8$B neutrino CE$\nu$NS without the DM, and b) the low mass DM with the $^8$B CEvNS background, the expected counts can be written as:
% \begin{equation*}
% \centering
%     \lambda_i  = ( 1+f_i - f_i \eta_{mod}) \cdot \eta_{cut} & \left[  \theta_{i,BDT_s} \cdot ( N_{wimp} + \eta_{flux} \cdot N_{\nu})  + \theta_{i,BDT_s} \cdot \theta_{i,AC} \cdot N_{AC} \right]] + N_{other},
%  \label{eq:lambda}
% \end{equation*}

\begin{equation}
\small
\begin{aligned}
\lambda^{\nu}_i  = & N_\nu (1+\delta_s f^\nu_i) (1+\delta_\epsilon) (1+\delta^i_{\textrm{BDT}, s}) + \\
& N_{\textrm{AC}} (1+\delta_b) (1+\delta_\epsilon) (1+\delta^i_{\textrm{BDT},b}) + N_{\textrm{other}}, \\
\lambda^{\chi}_i  = & N_\chi (1+\delta_s f^\chi_i) (1+\delta_\epsilon) (1+\delta^i_{\textrm{BDT}, s}) + \\
  & N_\nu (1+\delta_s f^\nu_i) (1+\delta_\epsilon) (1+\delta^i_{\textrm{BDT}, s}) (1+\delta_\Phi) + \\
& N_{\textrm{AC}} (1+\delta_b) (1+\delta_\epsilon) (1+\delta^i_{\textrm{BDT},b}) + N_{\textrm{other}},  \\
%\lambda^{\nu}_i  = & ( 1 + f^{\nu}_i\eta_{\textrm{mod}}) \eta_{\textrm{cut}}   \theta_{i,\textrm{BDT}_s}  \cdot N_{\nu} \\
% & + \eta_{\textrm{cut}}\cdot\theta_{i,\textrm{BDT}_{\textrm{AC}}} \cdot \eta_{\textrm{AC}} \cdot N_{\textrm{AC}}  + N_{\textrm{other}}, \\
\end{aligned}
   \label{eq:lambda}
\end{equation}
%where $\lambda^{\chi}_i$ and $\lambda^{\nu}_i$ are the expected count in the two hypotheses. 
where $N_\nu$, $N_{\textrm{AC}}$, $N_{\textrm{other}}$, and $N_{\chi}$ are the nominal numbers of counts for the $^8$B CE$\nu$NS, AC, other background events (including ER and neutron), and low mass DM, respectively.
% In the interpretation for $^8$B CE$\nu$NS events, the signal term for $N_{wimp}$ is removed, replaced by $^8$B CE$\nu$NS term but with scaling factor of flux $\eta_{flux}$ removed.
The total backgrounds predicted in the 2- and 3-hit ROI for the solar $^8$B neutrino search are 1.50 and 0.07, respectively, in an exposure of 0.48\,tonne$\cdot$year, as shown in Table~\ref{tab:bkg}.
% The probability of observing 3\,$\sigma$ excess is estimated to be xx\%.
% ROI of majority of data were blinded until the data selection, signal and background models ready.
% After unblinding, 1 and 0 events are found in 2- and 3-hit ROI that survive the selection.
The observed number of events is consistent with both background-only hypotheses in searching for the $^8$B CE$\nu$NS and the low mass DM in Eqn.~\ref{eq:lambda}, representing a probability of 53\% and 17\% of observing the same or less number of events than the data, respectively. 
% Their ($S2$, $S1$) and position distributions are shown in Fig.~\ref{fig:data_unblinded}.
% \textcolor{red}{
% After unblinding, we also found 1-hit region has significantly higher event count of 34 than our expected background of 8.78, which uses AC model based on fixed-window pairing.
% Further investigation shows that AC model based on in-order pairing gains much higher prediction of 28.7 which is more consistent with observation, indicating AC model after BDT applied in 1-hit region having large systematic uncertainty.
% However, this does not affect our results because discrepancy between AC models using in-order and fixed-window pairing are not observed in 2- and 3-hit regions.
% }

% Results
Using a similar procedure as in Refs.~\cite{meng2021dark, baxter2021recommended}, we give the 90\% C.L. upper limit on the solar $^8$B neutrino flux using the CE$\nu$NS channel, pushing the upper limit to 9.0$\times$10$^6$/cm$^2$/s, in comparison to  (5.46$\pm$ 0.66)$\times$10$^6$/cm$^2$/s from the standard solar model B16-GS98~\cite{vinyoles2017new}.
If the signal model adopted by XENON1T~\cite{aprile2021search} is used, the upper limit of the solar $^8$B neutrino flux will be lowered by 13\%.
If the signal model uncertainty ($\delta_s$) is eliminated from the fit, the upper limit will be reduced by 10\%.
Under the nominal $^8$B CE$\nu$NS rate, we also obtain the best constraints on the spin-independent DM-nucleon cross section with mass in the range of 3 to 9\,GeV/c$^2$. The results are summarized in Fig.~\ref{fig:results}.
In Fig. 4, we also show the $^8$B neutrino floor curves from Ref.~\cite{ruppin2014complementarity} under ideal background assumption. 
The current stage of PandaX has clearly entered into the sensitive region for neutrinos, so this result could also be cast into interesting parameter space of neutrino interactions. 
The lack of CE$\nu$NS excess from this work and XENON1T~\cite{aprile2021search} also motivates further investigations on the response of LXe TPC to ultralow energy nuclear recoils.

% Using such results, we give updated constraint on non-standard interaction between neutrinos and quarks, and on the elastic interaction between spin-independent low-mass WIMP and nucleon, which are shown in Fig.~\ref{fig:results1}.
% We follow the calculation in~\cite{} for non-standard neutrino interaction and in~\cite{} for low-mass dark matter.
% The most stringent constraints for low-mass WIMP in mass range from xx to xx\,GeV/c$^2$ are reached with lowered threshold in this analysis.
% Also we perform a study of active and sterile neutrino magnetic moment using these data.
% Following the calculation in~\cite{}, we obtained the best constraint on the active-sterile neutrino transition magnetic moment $\mu_{l4}$ in mass range from xx to xx\,MeV of sterile neutrino.
% Fig.~\ref{fig:results2} shows the constraint on $\mu_{l4}$ as a function of sterile neutrino mass $m_4$.

In summary, a search for CE$\nu$NS from the solar $^8$B neutrinos as well as the low mass DM-nucleon interactions is performed using the PandaX-4T commissioning data with 0.48\,tonne$\cdot$year exposure. In the analysis, we have further optimized the data selection and developed various techniques to lower the energy threshold and to control the accidental background. No significant excess is observed, leading to the strongest upper limit on the solar $^8$B neutrino flux using CE$\nu$NS, and on the spin-independent DM-nucleon cross section within the mass range from 3 to 9\,GeV/c$^2$.
This manifests the potential of PandaX-4T as a highly sensitive multi-purpose dark matter and astrophysical neutrino observatory.

% !TEX root = ../main.

% \section{Acknowledgement}

We would like to thank Matthew Szydagis for useful discussions concerning NEST model uncertainty.
This project is supported in part by grants from National Science
Foundation of China (Nos. 1209061, 12005131, 11905128, 11925502), a grant from the Ministry of Science and Technology of China (No. 2016YFA0400301),
and by Office of Science and
Technology, Shanghai Municipal Government (grant No. 18JC1410200). We thank supports from Double First Class Plan of
the Shanghai Jiao Tong University. We also thank the sponsorship from the Chinese Academy of Sciences Center for Excellence in Particle
Physics (CCEPP), Hongwen Foundation in Hong Kong, Tencent
Foundation in China and Yangyang Development Fund. Finally, we thank the CJPL administration and
the Yalong River Hydropower Development Company Ltd. for
indispensable logistical support and other help.
% \appendix
% \renewcommand{\appendixname}{Supplementary material~\Alph{section}}

% \section{Supplementary material: waveform simulation}
% \label{sec:appendix_a}
% \textcolor{red}{(This is just draft for content):}
% The statistics of calibration data in the ROI is limited mainly due to the emerged large amount of noise degrading the data quality.
\textbf{Appendix on waveform simulation.}
To have sufficient high-purity samples for estimating the efficiency in the ROI and for training the boosted decision tree (BDT) algorithm, a waveform simulation (WS) is developed, which includes our best knowledge from the data.
The WS not only simulates the $S1$ and $S2$ pulses, but also simulates the accompanying noises that could appear in the event waveform, such as the PMT afterpulsing, the delayed electrons, the photo-ionization, and the spurious $S1$s~\cite{sorensen2017two, sorensen2017electron,akerib2020investigation,akerib2021improving}.

\begin{figure}[htp]
    \centering
    \includegraphics[width=0.48\textwidth]{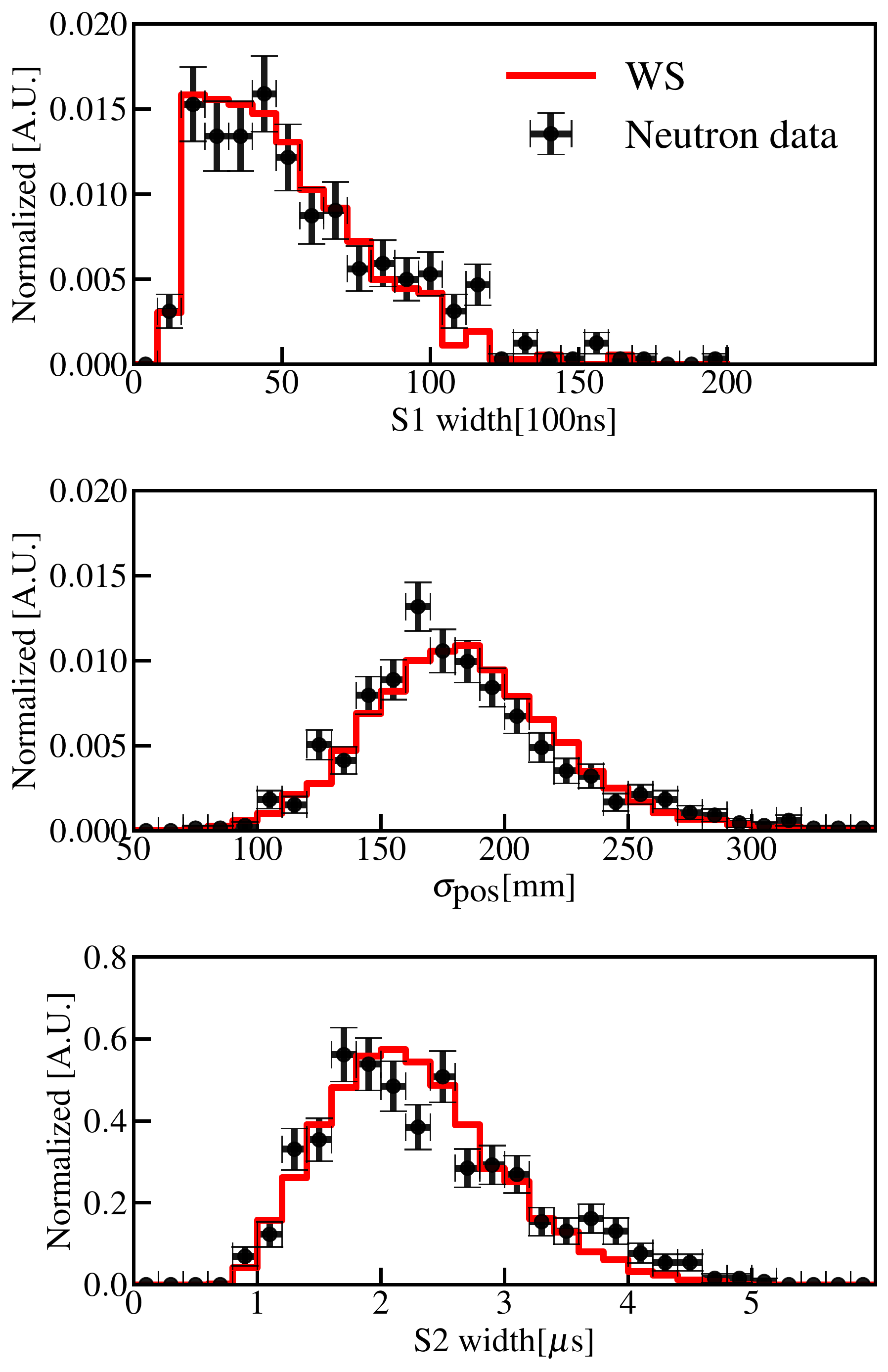}
    \caption{
    The comparisons between the neutron calibration data and WS of $S1$ width (top), $\sigma_{\textrm{pos}}$ (middle) and $S2$ width (bottom) normalized distributions in the ROI.}
    % \mwba{(recommend smaller title size to be the same as fig 3 and 4}}
    \label{fig:ws_width_comparison}
\end{figure}

% S1 and S2
The simulation of the $S1$ and $S2$ pulse waveforms is data driven. 
The simulated $S1$ pulse waveform is sampled using the real $S1$ hits from the neutron calibration data with the charge from 20 to 80\,PE.
%The $S1$s with the charge from \textcolor{red}{xx} to \textcolor{red}{xx} are selected, and xx\% to xx\% of their hits are randomly kept to construct the simulated $S1$s.
% The $S1$s with the charge from 20 to 80\,PE are selected, which are also required to have a consistent top-to-bottom ratio with that of the simulated $S1$.
% \textcolor{blue}{, and the charge fraction on the top and bottom array is tuned to match the light propagation process of data.}
The width distribution of the simulated $S1$s and $S1$s from the neutron calibration data in the ROI can be found in the top panel of Fig.~\ref{fig:ws_width_comparison}.
The simulated $S2$ pulse waveform is re-assembled using the single electron (SE) waveforms obtained from the data.
% The simulated $S2$ pattern is adjusted by effectively tuning the parameter $R_0$, which is the maximal distance between the reconstructed position of the selected SE event to the position of the simulated event.
The SEs are sampled within a circle with a radius $R_0$=40\,mm based on their reconstructed positions.
$R_0$ is tuned to match the root-mean-square distance of all fired top PMTs, weighted by charge, from the position of the top PMT that sees the most $S2$ charge ($\sigma_{\textrm{pos}}$).
% $R_0$ is tuned to match the average root mean square variance of the hit PMT positions of $S2$s ($\sigma_{\textrm{pos}}$) between the data and WS.
% $\sigma_{\textrm{pos}}$ is defined as 
% $\sqrt{\sum_i q_i (\textbf{x}_i-\hat{ \textbf{x}})^2 / \sum_i q_i}$
% $\sqrt{\frac{\sum^{top hits}[q_i((x_{PMTi}-x_{rec})^2+(y_{PMTi}-y_{rec})^2)]}{\sum^{top hits}q_i}}$
% where $\textbf{x}_i$, and $q_i$ are the hit PMT position and the charge, respectively, of the i-th $S2$ hit on top PMT array.
% $\hat{\textbf{x}}$ is the position of the top PMT that sees the most $S2$ charge.
With $R_0$=40\,mm, the comparison of $\sigma_{\textrm{pos}}$ between the neutron calibration data and WS are shown in the middle panel of Fig.~\ref{fig:ws_width_comparison}.
% The comparisons are shown in Fig.~\ref{fig:ws_width_comparison}.
% The mean time of the SE waveforms in the simulated event is sampled following 
The pileup of the SEs is required to follow a Gaussian distribution with the Gaussian $\sigma$ equals to $\sqrt{2 D T}$, where $D$ is the longitudinal diffusion coefficient in LXe and $T$ is the drift time of the simulated $S2$. 
The value of $D$ is obtained to be 28\,$\rm{cm}^2/s$ by matching the $S2$ width vs. drift time distribution of the neutron calibration data.
The comparison of the $S2$ width distribution between the neutron calibration data and WS in the ROI can be found in the bottom panel of Fig.~\ref{fig:ws_width_comparison}.

% \begin{figure}[htp]
%     \centering
%     \includegraphics[width=0.48\textwidth]{B8_PandaX_4T/plot/S2_AP.pdf}
%     \caption{
%     The probability distribution functions for the two components of delayed $S2$s: the one from the gate electrode (green solid line) and from the sensitive LXe volume (green solid line).
%     }
%     \label{fig:s2_ap}
% \end{figure}

\begin{figure}[htp]
    \centering
    \includegraphics[width=0.42\textwidth]{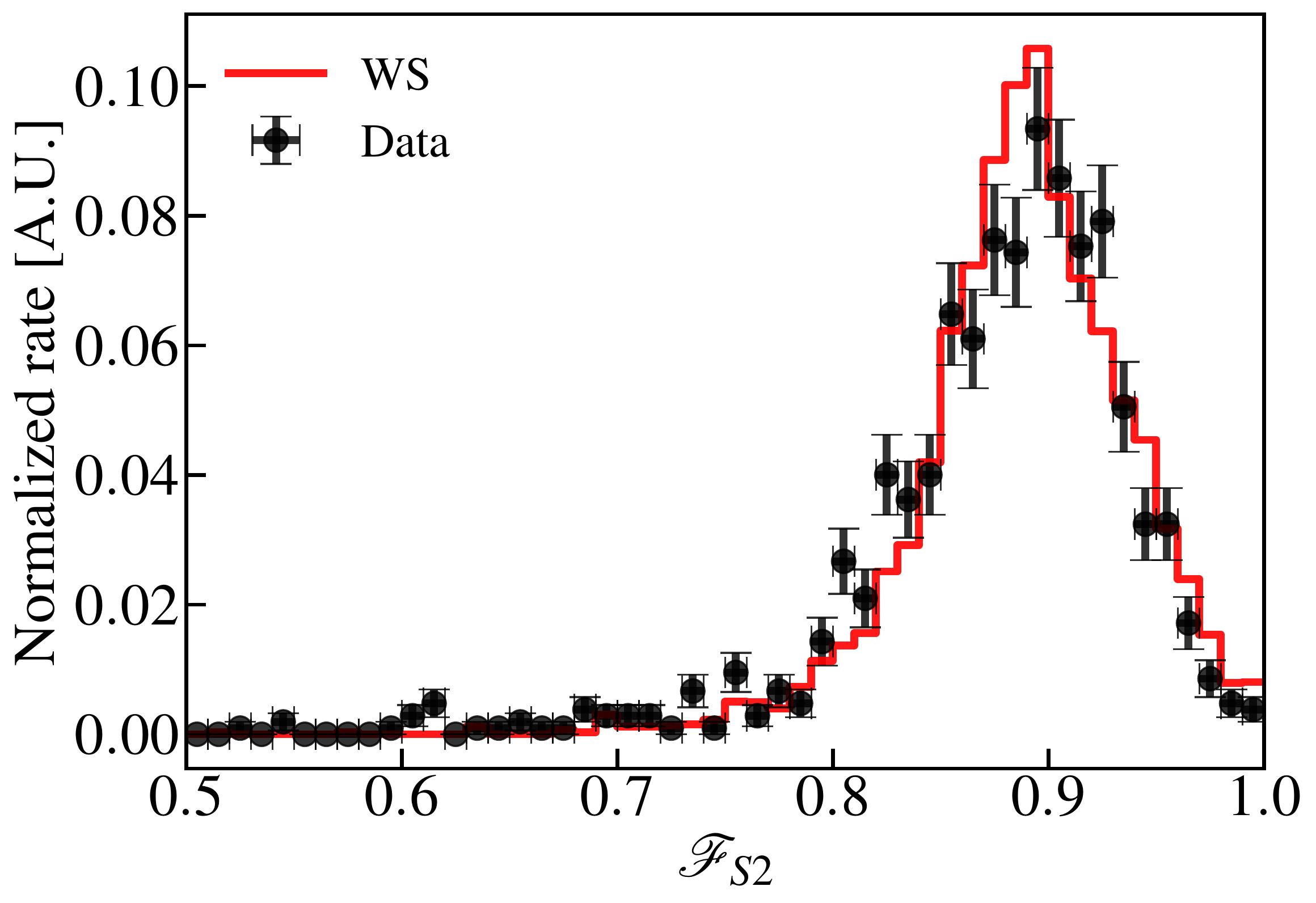}
    \caption{
    Comparison between the $\mathcal{F}_{S2}$ distributions from the WS (red solid line) and data (black dots with error bars).
    }
    \label{fig:qelse_comparison}
\end{figure}

% Noise
% old
% Dark counts and noises are included by inserting randomly picked 1-ms-long waveforms from all the recorded waveforms into the assembled $S1$ and $S2$ waveform.
% Other well-known processes, including PMT afterpulsing, delayed electrons, and electrons due to photo-ionization of the impurities, are taken into account in the WS.
% Since simulated $S1$ and $S2$ are both sampled using waveforms from the data, the PMT afterpulsing has already been included.
% The effects of delayed electrons and impurity photo-ionization can cause small $S2$ signals after a large $S2$.
% There are two primary components of the delayed $S2$s: the one originated from the gate electrode (approximately 3\,$\mu$s from the preceding $S2$) and from the LXe sensitive volume (more uniformly distributed in the sensitive volume).
% The time profile and probability of such delayed $S2$s are obtained by analyzing the data waveforms after the main $S2$.
% Fig.~\ref{fig:qelse_comparison} shows the comparison between $\mathcal{F}_{S2}$ distributions in events from the data and WS.

% \textcolor{violet}{
% (Rewrite para 3 (still feel inadequate and wordy):)
% }

% new
Dark counts and noises are included by inserting randomly picked 1-ms-long waveforms from all the recorded waveforms into the simulated event window.
PMT afterpulsing are already included, since the simulated $S1$ and $S2$ are both sampled using the waveforms from the data.
Delayed electrons and impurity photo-ionization can cause small delayed $S2$ signals after a large $S2$.
% For the delayed $S2$s, there are two primary components: one originated from the gate electrodes ($\sim$3\,$\mu$s from the preceding $S2$) and the other from the impurities or delayed electrons in the LXe (more uniformly distributed in time).
% There are two primary components of the delayed $S2$s: the one originated from the gate electrode (approximately 3\,$\mu$s from the preceding $S2$) and from the LXe sensitive volume (more uniformly distributed in the sensitive volume).
The time profile and probability of such delayed $S2$s are obtained by analyzing the data waveforms after the main $S2$.
The parameter which is mostly sensitive to the noise and afterglows is the $\mathcal{F}_{\textrm{S2}}$ that is defined in the main text.
% By combining all these effects together, $\mathcal{F}_{S2}$ is representative for checking the simulation result. % must have summary sencence
Fig.~\ref{fig:qelse_comparison} shows the comparison between the $\mathcal{F}_{S2}$ distributions from the neutron calibration data and the WS.

\end{document}